\documentclass{PoS}

\title{Polarized deuteron charge-exchange reaction
 ${dp\to \{pp\}_sN\pi}$ in the $\Delta$ isobar region}

\ShortTitle{Polarized deuteron charge-exchange}

\author{\speaker{Yu.N.~Uzikov}\\
Laboratory of Nuclear Problems, Joint Institute for Nuclear Research,
 RU-141980 Dubna, Russia \&
Department of Physics, Moscow State University, 119991 Moscow, Russia\\
        E-mail: \email{uzikov@jinr.ru}}

\author{J.~Haidenbauer \\
Institute for Advanced Simulation, Forschungszentrum J\"ulich,
D-52425 J\"ulich, Germany \\
        E-mail: \email{j.haidenbauer@fz-juelich.de}}

\author{C.~Wilkin \\
Physics and Astronomy Department, UCL, London, WC1E 6BT, UK\\
        E-mail: \email{c.wilkin@uc.ac.uk}}

\abstract{Mechanisms of the charge exchange reaction $dp\to \{pp\}_{\!s}
N\pi$, where $\{pp\}_{\!s}$ is a two-proton system at low excitation energy,
are studied at beam energies 1 -- 2~GeV and for invariant masses $M_X$ of
the final $N\pi $ system that correspond to the formation of the $\Delta(1232)$
isobar. The direct mechanism, where the initial proton is excited into the
$\Delta(1232)$, dominates and explains the existing data on the unpolarized
differential cross section and spherical tensor analyzing power $T_{22}$ for
$M_X> 1.2$~GeV/$c^2$. However, this model fails to describe $T_{20}$.}

\FullConference{XXII International Baldin Seminar on High Energy Physics Problems,\\
		15-20 September 2014\\
		JINR, Dubna, Russia}

\usepackage{graphics}
\usepackage{epsfig}
\newcommand{\pol}[1]{\mathaccent"017E{#1}}
\begin{document}


The $dp\to \{pp\}_{\!s}n$ reaction at low momentum transfers from the
incident deuteron to the final diproton $\{pp\}_{\!s}$ is sensitive to the
spin-flip part of the nucleon-nucleon charge-exchange
forces~\cite{buggwilkin}. Here $\{pp\}_{\!s}$ is a $pp$ pair at very low
excitation energy, typically $E_{pp}< 3$~MeV, where it is predominantly in
the $^{1\!}S_0$ state. A systematic study of this reaction has been started
at ANKE$@$COSY in both single~\cite{chiladze}  and
double-polarized~\cite{kacharava} experiments. In addition to the $pn\to np$
subprocess, there are variants of this reaction, namely $dp\to \{pp\}_{\!s}n
\pi^0$ or $dp\to \{pp\}_{\!s}p \pi^{-}$, that involve the spin-flip part of
the $pn\to \Delta^+(1232)n$ transition, which is difficult to measure
directly.

A linear combination of the Cartesian tensor analyzing powers $A_{xx}$ and
$A_{yy}$  in the $\pol{d} p \to \{pp\}_{\!s}\Delta^0(1232)$ reaction was
measured at SATURNE with a polarized deuteron beam of energy
2~GeV~\cite{ellegaard,elleg2} and a phenomenological analysis performed using
one-pion exchange~\cite{dmitrnpa}. New data on the unpolarized cross sections
and tensor analyzing powers of the $\pol{d} p\to \{pp\}N\pi$ reaction were
recently obtained at ANKE$@$COSY at energies 1.6, 1.8, and 2.3~GeV, where both
$A_{xx}$ and $A_{yy}$ were determined individually~\cite{d1d2,david2}.

All data on the $dp\to \{pp\}_{\!s} n$ reaction can be well explained by the
single-scattering mechanism with a $pn\to np$ sub-process, provided that one
accounts for the final $pp$ interaction in the $^{1\!}S_0$ state. It is
important to check whether the tensor analyzing powers of the $dp\to
\{pp\}_{\!s} N\pi$ reaction can be similarly described.

\begin{figure}[ht]
\begin{center}
\psfig{figure=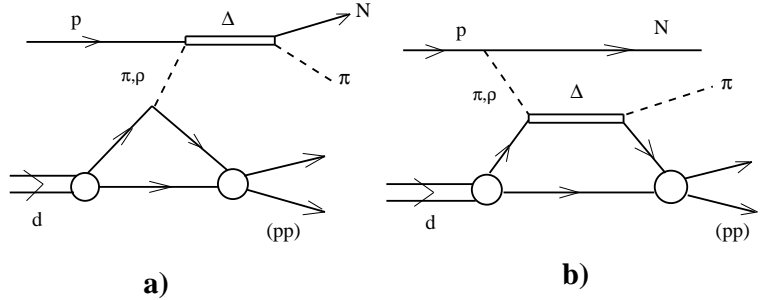,width=10.cm}
\end{center}
\vspace*{-0.5cm}
\caption{The  mechanisms of the
$dp\to \{pp\}_{\!s}N\pi$ reaction: a) direct (D), b) exchange (E).
\label{DEfig}
}
\end{figure}

It is expected that at low momentum transfers the $dp\to \{pp\}_{\!s} N\pi$
reaction is dominated by the direct (D) one-pion-exchange mechanism of
Fig.~\ref{DEfig}a. The differential cross section was evaluated within this
model~\cite{david2} using a modified form of the input employed for the
$p(^3\textrm{He},t)\Delta^{++}$ reaction~\cite{ellegaard}. At high $N\pi$ invariant
masses, $M_X\sim 1.2-1.35$~GeV/$c^2$, this explains well the shape of the measured
spectra, but it fails for lower masses~\cite{david2}. The E-mechanism of
Fig.~\ref{DEfig}b, where the $\Delta(1232)$ is excited in the deuteron, is of
little importance and its influence on the analyzing powers will be
neglected. For the elementary $pN\to\Delta N$ amplitudes we use both
$\rho$-meson and pion exchange.

We consider the mechanisms of Fig.~\ref{DEfig} on the basis of the Feynmann
diagram technique. For the meson-baryon vertices we apply the formalism used
in Ref.~\cite{imuz88}, where the exclusive $pp\to pn\pi^+$ data~\cite{lampf}
were analyzed in the $\Delta$-isobar region and the cut-off parameters at the
$\pi(\rho) NN$ and $\pi(\rho) N\Delta$ vertices were fixed from a fit to the
data.
The vertex
form factors $\pi(\rho) NN$ and $\pi(\rho) N\Delta$ are taken in the monopole
form, $F_{\pi(\rho)} (k^2)=(\Lambda^2-m_{\pi(\rho)}^2)/(\Lambda^2-k^2)$,
where $m_\pi$ $(m_{\rho})$ is the pion ($\rho$ meson) mass, $k$ is the pion
($\rho$ meson) 4-momentum, and $\Lambda$ is the cut-off parameter. The
$q^3$-dependence of the total width of the $\Delta$-isobar on the relative
momentum $q$ in $\pi N$ system is taken into account. The transition form
factor $d\to \{pp\}_{\!s}$ is evaluated using the CD-Bonn interaction
potential~\cite{CD-Bonn}.
\begin{figure}[ht]
\begin{center}
\psfig{figure=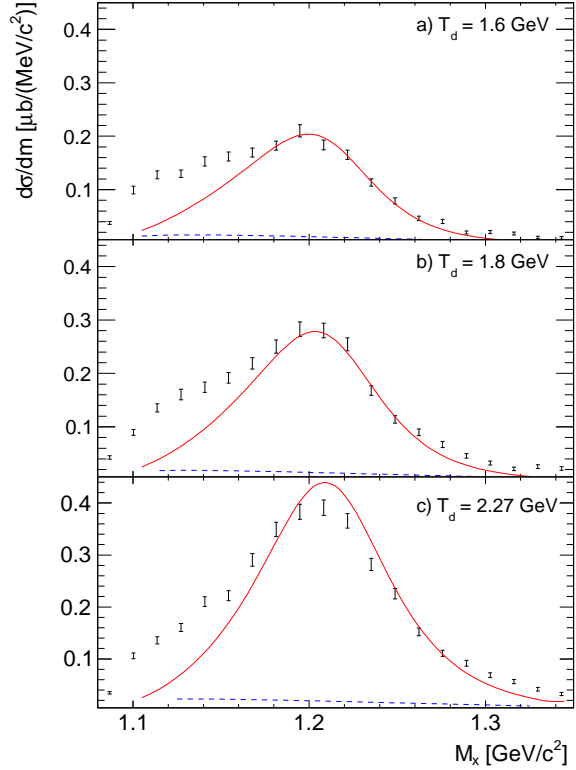,width=8cm}
\end{center}
\vspace*{-0.5cm}
\caption{The $dp\to \{pp\}_{\!s}N\pi$ differential cross section as a
function of the $\pi N$ invariant mass $M_X$ at three beam energies. The
ANKE@COSY data~\cite{david2} are compared with the calculations based on the
D- (full line) and E-mechanisms (dashed line) in impulse approximation. }
 \label{numbers}
\end{figure}

As shown in Fig.~\ref{numbers}, the D-mechanism can explain the shape of
$d\sigma/dM_X$ for $M_X>1.2$~GeV/$c^2$ at all three beam energies
studied~\cite{david2}. The magnitudes of the cross section are also
reasonably reproduced with a cutoff parameter value
$\Lambda=0.5$~GeV~\cite{imuz88}.

At lower masses, $M_X<1.2$~GeV/$c^2$, the D-mechanism fails to explain the
measured cross section~\cite{david2} and other mechanisms must be investigated.
The E-mechanism is calculated in a similar manner to the D. In this case, due
to spin-flip in the loop caused by the $d\to \{pp\}_{\!s}$ transition, the
vector product $[{\bf k}\times {\bf k}']$ of the momenta of pions appears in
the reaction amplitude. The E-contribution has indeed a maximum at low masses
$M_X\approx 1.1$~GeV/$c^2$. However, it is much smaller in absolute value
than the D-contribution (see dashed line in Fig.~\ref{numbers}) and therefore
does not provide an explanation of the observed shape of the cross section as
a function of $M_X$. The reasons for this small size are (i) the smallness of
the $\Delta-$propagator for the E-mechanism as compared to the D-mechanism,
and (ii) the smallness of the vector product $[{\bf k}\times {\bf k}']$ for
E-kinematics as compared  to the scalar product $({\bf k}\cdot {\bf k}')$ for
the D-kinematics.

As a check, we calculated the $dp\to dX$ cross section at almost the same
kinematics. The E-mechanism is here allowed but the D-mechanism is forbidden
by isospin. We found reasonable agreement with the experimental data on this
reaction~\cite{baldini} and also with the model calculations given in this
paper.

In impulse approximation the transition matrix element for the direct
mechanism of the $dp\to \{pp\}_{\!s}\Delta^0$ reaction can be written as
\begin{eqnarray}
M_{fi}= \Psi^+_j (\lambda_\Delta) (D_\pi k_j T_i +D_\rho M_{ji})\,
e_i(\lambda_d)\, \chi_p(\sigma_p),
\label{mfid}
\end{eqnarray}
where $\Psi^+_j$ is the vector-spinor of the $\Delta$-isobar, $\chi_p$ is the
spinor of the initial proton, $ e_i$ is the polarization vector of the
deuteron, $\lambda_\Delta$, $\lambda_d$ and $\sigma_p$ are spin-projections
of the $\Delta$, deuteron, and proton, respectively, and $k_j$ is the
3-momentum of the pion in the $\Delta$-isobar rest frame ($i,j=x,y,z$).
The factors $D_\pi$ and $D_\rho$ in
Eq.~(\ref{mfid})  are given by products of the coupling constants $\pi NN$,
$\rho NN$, $\pi N\Delta$, $\rho N\Delta$, form factors, and propagators of
the $\pi$ and $\rho$ mesons.
The vector operator for pion exchange $T_i$ is
 \begin{eqnarray}
T_i=\left(S_S(q)+\frac{1}{\sqrt{2}}S_D(q)\right)Q_i-
\frac{3}{\sqrt{2}}S_D(q)({\bf Q}\cdot {\bf n})n_i,
\label{tfipi}
\end{eqnarray}
where $S_S(q)$ and $S_D(q)$ are the $S$- and $D$-wave $d\to \{pp\}_{\!s}$
transition form factors at 3-momentum transfer ${\bf q}$, and ${\bf n}$ is the
unit vector along ${\bf q}$. The momentum $\bf Q$ is
\begin{equation} \label{Qpn}
{\bf Q}=\left [\frac{E_1+m_N}{E_2+m_N}\right ]^{1/2}\!{\bf p}_2-
\left [\frac {E_2+m_N}{E_1+m_N}\right ]^{1/2}\!{\bf p}_1 ,
\end{equation}
where $E_i=\sqrt{p_i^2+m_N^2}$ and ${\bf p}_1$ $({\bf p}_2)$  is the 3-momentum
of the virtual proton (neutron).

The tensor $\mathcal{M}_{ji}$  describes $\rho$-meson exchange:
 \begin{equation}
\mathcal{M}_{ji}=(S_S+\frac{1}{\sqrt{2}}S_D)\bigl [
({\bf Q}\cdot{\bf Q}^\prime) \delta_{ji}-Q_j Q^\prime_i\bigr]
-\frac{3}{\sqrt{2}}S_D \bigl [({\bf Q}\cdot{\bf Q}^\prime)n_j
-({\bf Q}^\prime\cdot {\bf n})Q_j\bigr ]n_i,
\label{mrhoji}
\end{equation}
where ${\bf Q}^\prime$ is the momentum of the $\rho$ meson in the
$\Delta$-isobar rest frame.

\begin{figure}[ht]
\begin{center}
\psfig{figure=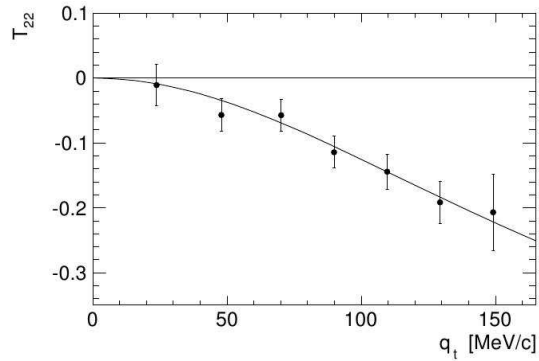,width=7.5cm}
\end{center}
\vspace*{-0.5cm}
\caption{Spherical tensor analyzing power $T_{22}=(A_{xx}-A_{yy})/2\sqrt{3}$ for the
$dp\to \{pp\}_{\!s}N\pi$ reaction averaged over the beam energies of
Ref.~\cite{david2}, versus the transverse momentum transfer $q_t$. }
 \label{figt22}
\end{figure}

The Cartesian tensor analyzing powers $A_{ij}$ are defined by $A_{ij}= Tr\{{\mathcal{M} \hat{\cal{P}}_{ij}
\mathcal{M}^+}\}/Tr\{\mathcal{M}\mathcal{M}^+\},$ where $\mathcal{M}$ is the
transition operator given by Eq.~(\ref{mfid}),
$\hat{\cal{P}}_{ij}=\frac{3}{2}(S_iS_j+S_jS_i)-\delta_{ij}$, and $S_l$ is the
spin-1 operator ($i,j,l=x,y,z$). Following the presentation of the ANKE@COSY
experiment~\cite{david2}, we consider the Cartesian tensor analyzing powers
$A_{xx}$ and $A_{yy}$ as functions of the transverse component of the
momentum transfer $q_t$. The $z$ axis is chosen to lie along the deuteron beam
momentum ${\bf p}_d$, $y$ along ${\bf p}_d\times {\bf p}_{pp}$,
where ${\bf p}_d$ (${\bf p}_{pp}$) is the total momentum of the deuteron ($pp$-pair),
 with $x$ being
taken so as to form a right-handed  coordinate system. The experimental data
of Ref.~\cite{david2} were summed over the invariant mass of the  undetected
$\pi+N$ system in the interval $1.19 < M_X < 1.35$~GeV/$c^2$ at fixed $q_t$.
Thus, we obtain, for example,
\begin{equation}
A_{xx}=1- 3\left.\int_{M_X^{min}}^{M_X^{max}} \mathcal{M}_{\alpha
x}\mathcal{M}^+_{\alpha x}dM_X\right/ \int_{M_X^{min}}^{M_X^{max}}
\mathcal{M}_{\alpha i}\mathcal{M}^+_ {\alpha i}dM_X\cdot \label{axx}
\end{equation}

Non-relativistically ${\bf Q}= {\bf p}_p-{\bf p}_n={\bf q}$.  In this limit,
and ignoring integration over $M_X$, one finds from Eq.~(\ref{axx}) that, for
pure $\pi$ and $\rho$ exchange,
\begin{eqnarray}
\nonumber
A_{xx}^{\pi}=1- 3{q_t^2}/{{\bf q}^2}\ &\textrm{and}&\ A_{yy}^{\pi}=1,\\
A_{xx}^{\rho}=-\textstyle{\frac{1}{2}}+ 3{q_t^2}/{2{\bf q}^2}\ &\textrm{and}&\
A_{yy}^{\rho}=-\textstyle{\frac{1}{2}}.
\label{axxpi}
\end{eqnarray}

Both of these simple limits are in contradiction with experiment, for which
$A_{xx}(q_t=0)=A_{yy}(q_t=0)\approx 0$ and $A_{yy}(q_t)$ has a smooth $q_t$
dependence~\cite{david2}. Calculations for $\pi+\rho$ exchange have been
performed with $\Lambda_{\pi NN}=\Lambda_{\pi N\Delta}=0.5$~GeV and
$\Lambda_{\rho NN}=\Lambda_{\rho N\Delta}=0.7$~GeV~\cite{imuz88}, when
$\rho$-exchange is almost negligible. For these parameters, $A_{xx}$
decreases with increasing $q_t$ from $A_{xx}=1$ at $q_t=0$ to $A_{xx}\approx
0.5$ at
$q_t=170$ MeV/c,
 whereas $A_{yy}$ is almost independent of $q_t$ and
close to unity.

When the contribution of the $\rho$ meson is raised by increasing the cut-off
parameter $\Lambda_{\rho NN}=\Lambda_{\rho N\Delta}$ to 1.3~GeV, both
$A_{xx}$ and $A_{yy}$ decrease but stay far from experiment at $q_t=0$. The
$\pi+\rho$ model also fails to describe the spherical tensor analyzing power,
$T_{20}=-(A_{xx}+A_{yy})/\sqrt{2}$. On the other hand, the spherical analyzing power
$T_{22}=(A_{xx}-A_{yy})/2\sqrt{3}$ is well described by the direct one-pion
exchange of Fig.\ref{DEfig}a, as demonstrated in Fig.~\ref{figt22}.

To take into account an interference  between the D and E mechanisms would
require one to consider the quasi-three-body final states $\pi^0 n
\{pp\}_{\!s}$ and $\pi^- p \{pp\}_{\!s}$  explicitly instead of the
quasi-two-body state $\Delta^0 \{pp\}_{\!s}$. However, this will not improve
the results at lower masses $M_X<1.2$~GeV/$c^2$. Our results suggest that
one should  use a $NN\to N\Delta$ amplitude beyond undistorted $\pi+\rho$
exchange.

The authors are grateful to A.~Kacharava and D.~Mchedlishvili for useful
discussions. This work was supported in part by the Heisenberg--Landau
programme.

\end{document}